\documentclass[preprints,article,accept,pdftex,moreauthors]{mdpi} 
 \firstpage{1} 
\pubvolume{16}
\issuenum{2}
\articlenumber{667}
\pubyear{2023}
\copyrightyear{2023}
\datereceived{13 December 2022} 
 \daterevised{4 January 2023} 
\dateaccepted{6 January 2023} 
\datepublished{10 January 2023} 
\hreflink{https://\linebreak doi.org/10.3390/ma16020667}
\Title{Nanocrystalline CaWO$_4$ and ZnWO$_4$ Tungstates for Hybrid Organic--Inorganic X-ray Detectors} %Attention: Title altered.

\TitleCitation{Nanocrystalline CaWO$_4$ and ZnWO$_4$ Tungstates for Hybrid Organic--Inorganic X-ray Detectors}

\Author{Inga Pudza $^{1,}$*\orcidA{}, Kaspars Pudzs $^{1}$\orcidB{}, Andrejs Tokmakovs $^{1}$\orcidC{}, Normunds Ralfs Strautnieks $^{1}$, \mbox{Aleksandr Kalinko $^{1,2}$\orcidD{}} and Alexei Kuzmin $^{1}$\orcidE{}}

\AuthorNames{Inga Pudza, Kaspars Pudzs, Andrejs Tokmakovs, Normunds Ralfs Strautnieks, Aleksandr Kalinko and Alexei Kuzmin }

\AuthorCitation{Pudza, I.; Pudzs, K.; Tokmakovs, A.; Strautnieks, N.R.; Kalinko, A.; Kuzmin, A.}
\address{%
	$^{1}$ \quad Institute of Solid State Physics, University of Latvia, Kengaraga Street 8, LV-1063 Riga, Latvia%; kaspars.pudzs@cfi.lu.lv (K.P.); 
	\\
	$^{2}$ \quad Deutsches Elektronen-Synchrotron DESY, Notkestrasse 85, D-22607 Hamburg, Germany}

\corres{Correspondence: inga.pudza@cfi.lu.lv}

\usepackage{amssymb}
\usepackage{amsmath}
\usepackage{gensymb}
\usepackage{textcomp}
\usepackage[version=4]{mhchem}
\usepackage{siunitx}

\let\Delta\varDelta
\let\Theta\varTheta
\let\Sigma\varSigma
\let\phi\varphi

\newcommand{\doHMN}[2]{%
	\begingroup\lccode`~=`#1
	\lowercase{\endgroup\let~}#2%
	\mathcode`#1="8000
}

\abstract{Hybrid materials combining an organic matrix and high-Z nanomaterials show potential for applications in radiation detection, allowing unprecedented device architectures and functionality. Herein, novel hybrid organic--inorganic systems were produced using a mixture of tungstate (CaWO$_4$ or ZnWO$_4$) nanoparticles with a P3HT:PCBM blend. The nano-tungstates with a crystallite size of 43~nm for CaWO$_4$ and 30~nm for ZnWO$_4$ were synthesized by the hydrothermal method. Their structure and morphology were characterized by X-ray diffraction and scanning electron microscopy. The hybrid systems were used to fabricate direct conversion X-ray detectors able to operate with zero bias voltage. The detector performance was tested in a wide energy range using monochromatic synchrotron radiation. The addition of nanoparticles with high-Z elements improved the detector response to X-ray radiation compared with that of a pure organic P3HT:PCBM bulk heterojunction cell. The high dynamic range of our detector allows for recording X-ray absorption spectra, including the fine X-ray absorption structure located beyond the absorption edge. The obtained results suggest that nanocrystalline tungstates are promising candidates for application in direct organic--inorganic X-ray detectors.}


\begin{document}

%%%%%%%%%%%%%%%%%%%%%%%%%%%%%%%%%%%%%%%%%%

\section{Introduction}
Nowadays, developing new radiation detectors based on nanomaterials is an active field of research~\cite{Luo2017}. Among~the different types of detectors, hybrid organic--inorganic systems for X-ray detection have attracted considerable attention during the last ten years~\cite{Thirimanne2018}. The~strong advantages of such systems are the combination of relatively inexpensive, easy-to-manufacture, flexible, and~low-bias voltage ($<$10~V) organic semiconductors with high-Z inorganic (nano-)compounds~\cite{Luo2017,Fraboni2012,Suzuki2014,Basirico2016}. The~latter provide a large X-ray cross-section and control of spectral selectivity, which improve the absorption efficiency and sensitivity while maintaining the beneficial physical properties of the host organic matrix~\cite{Thirimanne2018}. 
Various inorganic materials have been proposed for use in hybrid organic--inorganic systems for X-ray detection~\cite{Thirimanne2018}. However, the~search for the best system(s) that can be used in everyday applications remains one of the most important and challenging tasks in the field~\cite{Liu2019,Mescher2020}.

Here, we propose tungstates with a general chemical formula of AWO$_4$ \cite{Sleight1972} (where A is a divalent ion, for~example, Ca, Sr, Ba, Pb, Ni, Zn, or Cd) as a new class of materials for use in hybrid organic--inorganic systems for direct-conversion X-ray detection. Using tungstates opens up a wide range of possibilities for solving a specific problem by optimizing their chemical composition and degree of crystallinity. This is convenient for producing hybrid systems but has not been exploited until now. Another advantage of tungstates for use in X-ray detectors is the high Z of tungsten ($Z$ = 74) and the possibility of selecting the Z number of the second metal ion in a wide range. This allows for optimizing the absorption efficiency in a certain range of X-ray~energies. 

In this study, two possible candidates for use as the X-ray absorber in hybrid organic--inorganic direct X-ray detectors, i.e., nanocrystalline tungstates CaWO$_4$ and ZnWO$_4$ with the scheelite and wolframite crystallographic structures~\cite{Sleight1972}, respectively, were synthesized and characterized by X-ray diffraction (XRD) and scanning electron microscopy (SEM). The~X-ray detectors were fabricated based on a mixture of nanotungstates with a P3HT:PCBM blend, and~their ability to detect X-rays was demonstrated using tunable synchrotron~radiation.

\section{Materials and~Methods}
\unskip
\subsection{Nanoparticle Synthesis and~Characterization}

\textls[-25]{CaWO$_4$ and ZnWO$_4$ nanoparticles (NPs) were produced by the hydrothermal method~\cite{Lee2016}.} Citric acid C$_6$H$_8$O$_7$ was used as a surfactant/capping agent and provided kinetic grain size control~\cite{Su2007}. 

\begin{figure}[H]
	 	
	\includegraphics[width=1\textwidth]{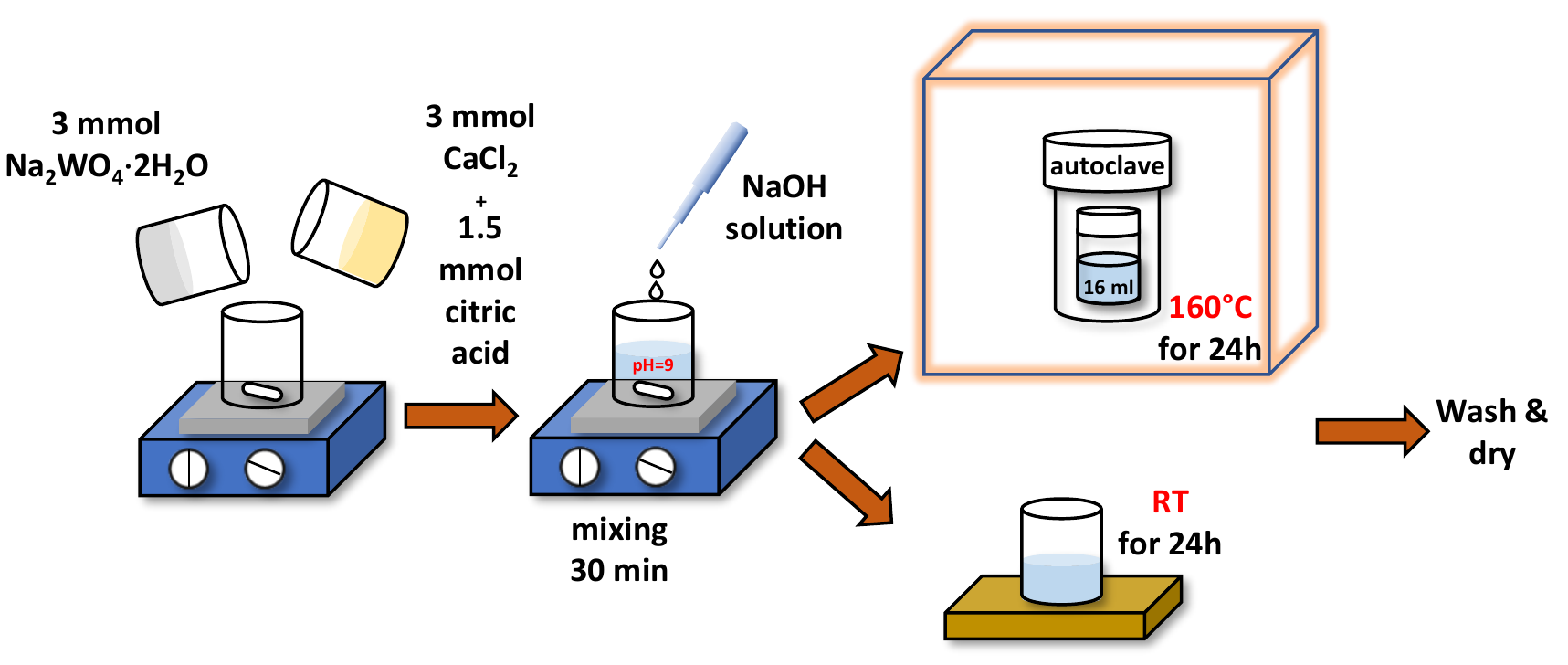}	
	\caption{Schematics of CaWO$_4$ nanoparticle synthesis. See text for~details. }
	\label{fig1}
\end{figure}

First, 3~mmol of CaCl$_2$ (97\%, Alfa Aesar, Haverhill, MA, USA) 
and Na$_2$WO$_4$$\cdot$2H$_2$O ($\ge$99\%, 97\%, Alfa Aesar, Haverhill, MA, USA) were separately dissolved in deionized water (Figure\ \ref{fig1}). Next,~citric acid (1.5~mmol) was added to the CaCl$_2$ solution, and~the obtained mixed solution was subsequently added to the Na$_2$WO$_4$$\cdot$2H$_2$O solution. 
The solution pH was adjusted to 9 by adding an appropriate amount of NaOH ($\ge$98\%, Sigma-Aldrich, St. Louis, MO, USA) solution in water. The~obtained solution was mixed under constant magnetic stirring for 30~min. One part (16~mL) of the resulting solution was sealed in a Teflon-lined stainless-steel autoclave (25~mL) and was allowed to react at $\sim$160~$\degree$C for 24~h, followed by natural cooling to room temperature. The~rest of the solution was left at room temperature (RT) for 24~h for a~comparison.

\vspace{-6pt}

The synthesis of ZnWO$_4$ NPs was analogous, only we used Zn(NO$_3$)$_2$$\cdot$6H$_2$O (98\%, Sigma-Aldrich, St. Louis, MO, USA) as the zinc source. Note that in this case, a~lower pH value (pH = 8) was used, because an increase in pH to 9 resulted in ZnO impurities. The~precipitates of nanotungstates obtained after hydrothermal treatment were washed and centrifuged several times in the following sequence: distilled water, isopropanol, acetone, and~chlorobenzene, and~finally dried in air at 75~$\degree$C.

The phase composition and crystallinity of all samples were controlled by powder X-ray diffraction (XRD) at room temperature using a benchtop Rigaku MiniFlex 600 diffractometer (Rigaku, Tokyo, Japan) with Bragg--Brentano geometry (Cu K$_\alpha$ radiation), operated at {40}{ kV} and {15}~{mA}. The~crystallite size of NPs was estimated using the Rietveld refinement method as implemented in Profex software~\cite{Doebelin2015profex}.

The sample morphology was studied by scanning electron microscopy (SEM) in immersion mode using a Helios 5 UX microscope (Thermo Fisher Scientific, Waltham, MA, USA) (Elstar in-lens SE TLD detector) operated at 2.00~kV. The~particle-size distribution was evaluated considering the statistics of 200 NP measurements in the SEM~micrographs.

\subsection{Hybrid Organic--Inorganic X-ray Detector Fabrication and~Measurements}

Hybrid organic--inorganic X-ray detectors were fabricated on top of the 25 $\times$ 25~mm ITO (In$_2$O$_3$:Sn)-coated glass with a sheet resistance of 5~$\Omega/sq$ (Pr\"azisions Glas \& Optik GmbH ). A poly(3,4-ethylenedioxythiophene)-poly(styrenesulfonate) (PEDOT:PSS; Heraeus Al4083) layer with a thickness of 40~nm was used as the hole transport and electron blocking layer. It was spin-coated in air (2500~rpm for 40~s with an acceleration of 2500~rpm/s) and annealed at 150~$\degree$C for 10~min.

A suspension of tungstate NPs and P3HT:PCBM was prepared by mixing the tungstate powder with a solution of the P3HT:PCBM mixture (weight ratio 1:1) in chlorobenzene (99.8\% anhydrous, Sigma-Aldrich, St. Louis, MI, USA), followed by sonication of the premix for 1 h. The~weight ratio of NPs:P3HT:PCBM in the suspension was 2:1:1. Thin films were fabricated from the suspension by the blade-casting method on a substrate heated to 75~$\degree$C. The~P3HT:PCBM mixture was crystallized by annealing at 140~$\degree$C for 15~min.

We deposited a 5~nm thick hole-blocking layer of 4,7-diphenyl-1,10-phenanthroline (BPhen, Sigma-Aldrich, St. Louis, MI, USA) and 100~nm thick Al electrode on top of the hybrid layer by thermal evaporation in a vacuum at a pressure of less than 7 $\times$ 10$^{-6}$~mbar. Al electrodes were deposited in a way that six separate ``active pixels'' with a size of \linebreak 4 $\times$ 4~mm were formed and could be independently tested. Thus, the~final hybrid organic--inorganic X-ray detector was composed of five ITO/PEDOT:PSS/NPs:P3HT:PCBM/BPhen /Al layers. The~detectors were additionally encapsulated with glass to reduce their possible degradation in air. No significant degradation of the detectors was observed during the~experiments.

X-ray measurements were conducted with the DESY PETRA-III P64 Advanced X-ray Absorption Spectroscopy undulator beamline~\cite{PETRAP64}. The~PETRA-III storage ring operated at $E$ = {6} {GeV} and current $I$ = {100}{ mA} in top-up 40 bunch mode. Fixed-exit double-crystal monochromator Si (111) was used to select the required X-ray energy from the undulator photon source. The~X-ray intensity $I_0$ before the sample was monitored with an ionization chamber. The~beam size on the sample was about 1 $\times$ 1~mm. The~hybrid organic--inorganic X-ray detector was
placed inside the vacuum chamber and connected to a Keithley 428 current amplifier. The~signal $I_{detector}$ was measured with the P64 beamline ionization chamber monitoring setup. The~absolute values of the detected photocurrent were in the range of 0.1--1.0~nA.
A passivated implanted planar silicon (PIPS) detector (Canberra) was used for the simultaneous detection of X-ray fluorescence. All measurements were recorded in a dark environment to exclude possible photoelectric effects induced by light in the experimental~hutch.

\section{Results and~Discussion}

\begin{figure}[H]
	 	
	\includegraphics[width=0.9\textwidth]{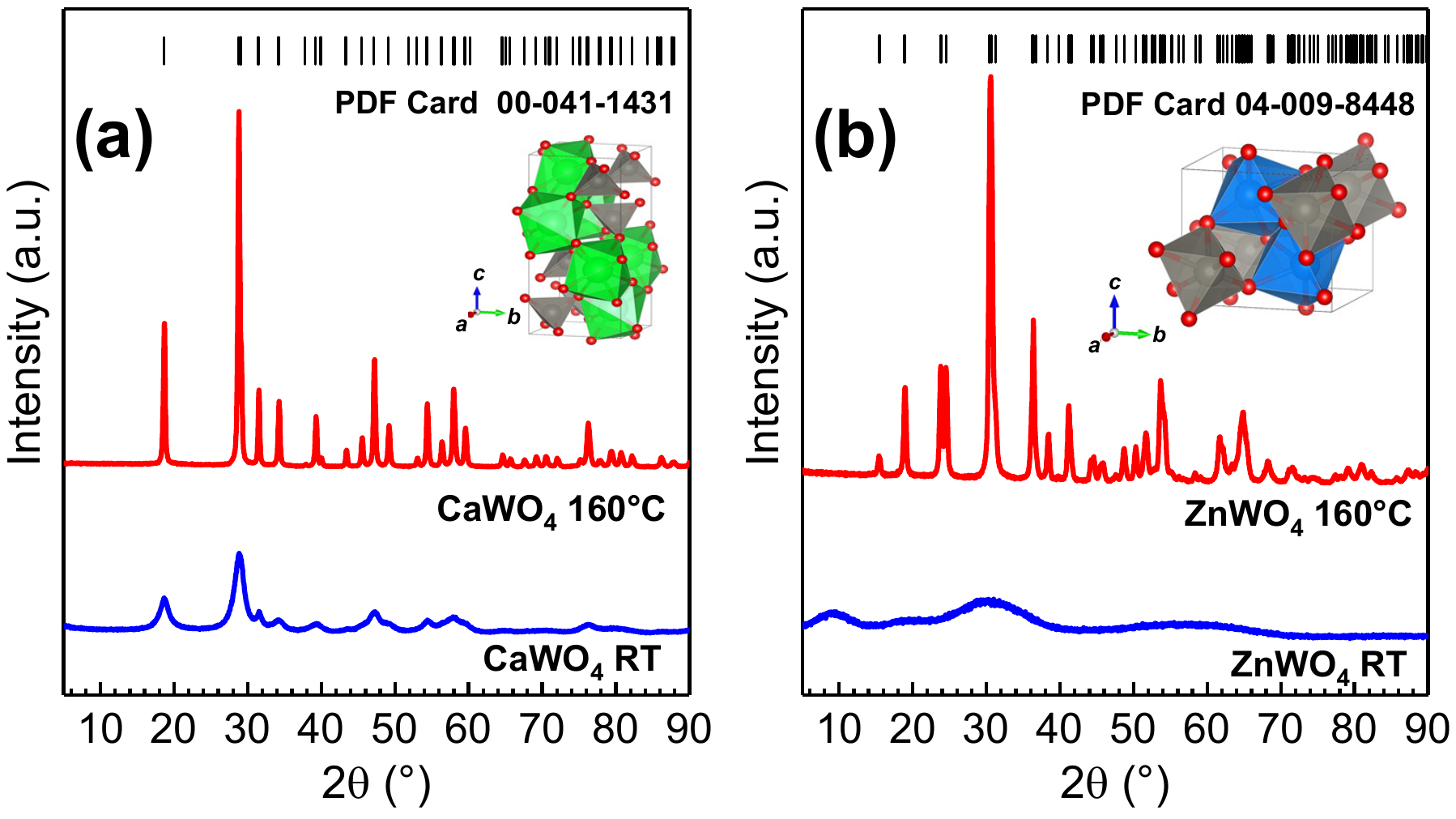}	
	\caption{X-ray diffraction patterns of CaWO$_4$ (\textbf{a}) and ZnWO$_4$ (\textbf{b}) nanoparticles left at RT (\textbf{bottom}) and treated at $\sim$160~$\degree$C for 24~h (\textbf{middle}). The~standard PDF cards of the CaWO$_4$ (PDF Card 00-041-1431) and ZnWO$_4$ (PDF Card 04-009-8448) phases are shown for comparison at the (\textbf{top}). The~crystal structures of the tungstates are shown in the~insets. }
	\label{fig2}
\end{figure}

Bulk CaWO$_4$ and ZnWO$_4$ tungstates have different crystallographic structures (scheelite and wolframite~\cite{Sleight1972}) composed of WO$_4$ tetrahedral and WO$_6$ octahedral units (see the insets in Figure\ \ref{fig2}), respectively. Both calcium and zinc ions are coordinated with oxygen ions; however, calcium ions have eight-fold coordination, while zinc ions have six-fold coordination. In~scheelite CaWO$_4$, Ca$^{2+}$ cations are located between slightly distorted WO$_4$ tetrahedra~\cite{Gurmen1971}, whereas in wolframite ZnWO$_4$, distorted WO$_6$ and ZnO$_6$ octahedra are connected by edges and form infinite zigzag chains~\cite{Schofield1996}. Both tungstates can be prepared in the nanocrystalline form~\cite{Liu2011,Pereira2018}.

The XRD patterns of CaWO$_4$ and ZnWO$_4$ nanocrystalline samples synthesized at RT and using hydrothermal treatment at $\sim$160~$\degree$C for 24~h in the autoclave are shown in Figure \ref{fig2}. The~CaWO$_4$ nanoparticles prepared at RT demonstrated weak crystallinity, while the ZnWO$_4$ nanoparticles were amorphous. At~the same time,
the XRD patterns of hydrothermally treated tungstate samples contained many Bragg peaks, which could be indexed to the pure tungstate phases (PDF Card 00-041-1431 for CaWO$_4$ and PDF Card 04-009-8448 for ZnWO$_4$): tetragonal phase (space group $I4_1/a$) with the lattice constants \linebreak $a$ = $b$ = 5.246~\AA\, and $c$ = 11.380~\AA\ for CaWO$_4$ and monoclinic phase (space group $P2/c$) with the lattice constants $a$ = 4.690~\AA, $b$ = 5.734~\AA\, and $c$ = 4.940~\AA\ for ZnWO$_4$. The~average crystallite sizes of hydrothermally grown CaWO$_4$
and ZnWO$_4$ nanotungstates were determined using Rietveld refinement and were equal to $\sim$43~nm and $\sim$30~nm, respectively. The~CaWO$_4$ seeds for nanoparticle synthesis left at RT for 24~h had a size of $\sim$5~nm. 

\begin{figure}[H]
	 
	\includegraphics[width=0.95\textwidth]{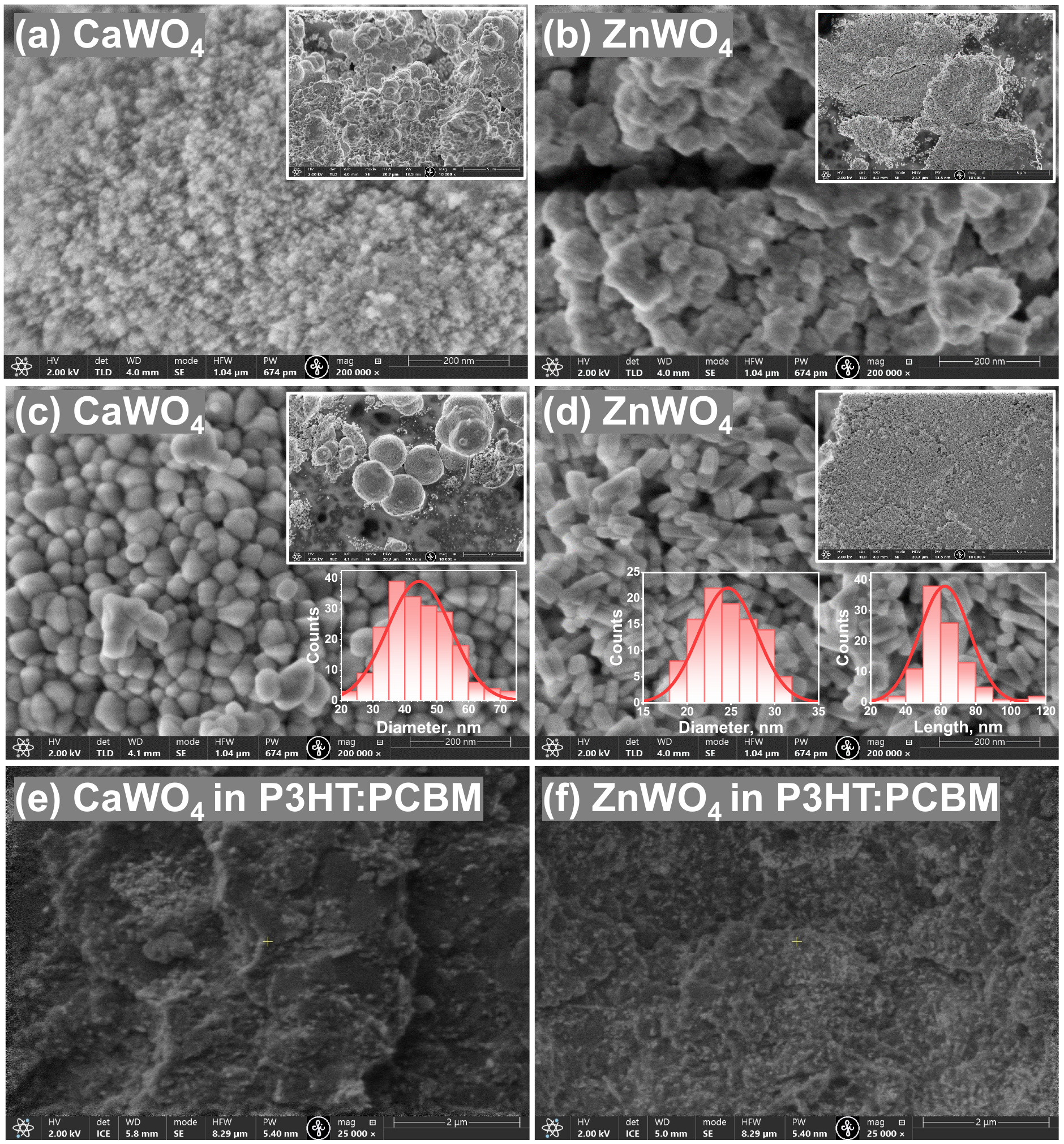}
	\caption{SEM micrographs of CaWO$_4$ and ZnWO$_4$ nanoparticles at RT (\textbf{a},\textbf{b}) and treated at $\sim$160~$\degree$C for 24~h (\textbf{c},\textbf{d}). 
		Cross-section SEM images of fabricated hybrid detectors (part of tungstate NPs incorporated in a P3HT:PCBM matrix) are also shown in (\textbf{e},\textbf{f}). 
		The particle-size distributions of CaWO$_4$ (diameter of nanoparticles) and ZnWO$_4$ (diameter and length of nanorods) estimated from the corresponding SEM micrographs are shown in the insets in (\textbf{c},\textbf{d}), respectively.}
	\label{fig3}
\end{figure}

\textls[-10]{Both tungstates have different morphology, as shown in the SEM micrographs \mbox{(Figure\ \ref{fig3})}. Products synthesized at RT formed fine powders with agglomerated particles (\mbox{Figure\ \ref{fig3}a,b)}. The~crystallinity was significantly improved after hydrothermal treatment at $\sim$160~$\degree$C as evidenced by the well-defined facets of the particles (Figure\ \ref{fig3}c,d). CaWO$_4$ formed microspheres composed of irregular polyhedral NPs having an approximate average diameter of 45~nm with a standard deviation of 10~nm (Figure\ \ref{fig3}c). This estimate is in agreement with the average crystallite size found from the XRD data. For~ZnWO$_4$, rod-like morphology and higher uniformity were evident (Figure\ \ref{fig3}d). Individual ZnWO$_4$ nanorods with a length of 62 $\pm$ 14~nm and a diameter of 25 $\pm$ 3~nm were resolved. The~average crystallite size estimated by XRD was between the determined length and diameter values of particles from the SEM~micrographs.}

%\unskip

Hydrothermally treated tungstate NPs were incorporated into a P3HT:PCBM matrix to fabricate a hybrid organic--inorganic direct-conversion X-ray detector. The~concentration of the NPs:P3HT:PCBM suspension used for the active layer deposition in this study was fixed at a weight ratio of 2:1:1. It can be seen in the cross-sectional SEM images (Figure\ \ref{fig3}e,f) that after mixing, the~NPs distributed fairly uniformly in the hybrid layer. The~thickness of the active layer with CaWO$_4$:P3HT:PCBM (ZnWO$_4$:P3HT:PCBM) was $\sim$17~\textmu m ($\sim$16~\textmu m).

\begin{figure}[H]
	 
	\includegraphics[width=0.98\textwidth]{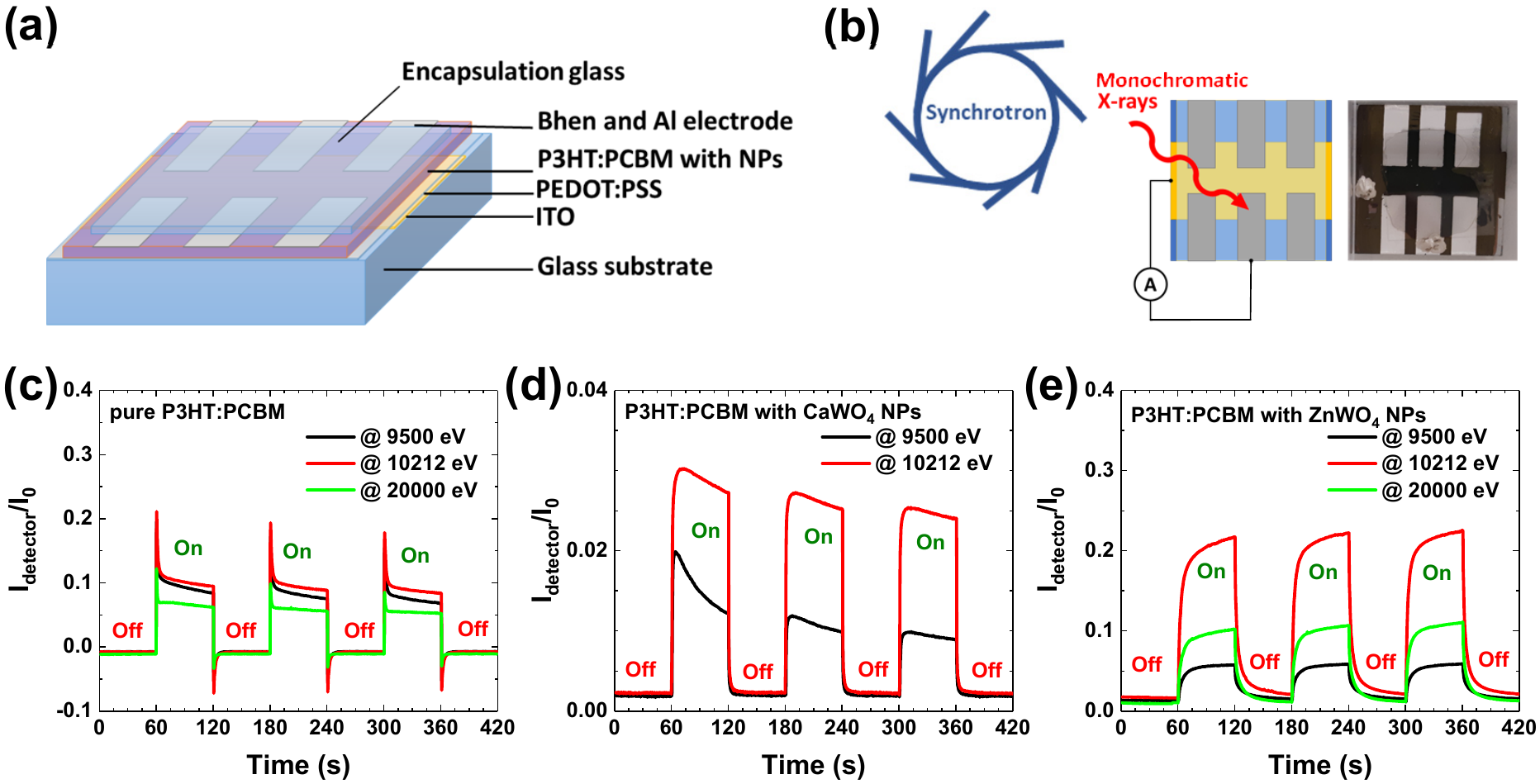}	
	\caption{A schematic representation of fabricated hybrid organic--inorganic direct-conversion X-ray detectors (\textbf{a}). A~scheme of X-ray response measurements at synchrotron, and a photo of fabricated X-ray detector (\textbf{b}). X-ray$-$induced response signals (I$_{detector}$/I$_{0}$) for pure P3HT:PCBM (\textbf{c}), hybrid CaWO$_4$+P3HT:PCBM (\textbf{d}), and~ZnWO$_4$+P3HT:PCBM (\textbf{e}) systems under repeated X-ray exposures (on/off cycles). } 
	\label{fig4}
\end{figure}

The X-ray detectors were fabricated with a sandwich-type architecture and multilayer stacking as depicted in Figure\ \ref{fig4}a. The~detectors were realized on the top of the ITO-coated glass substrate. A~polymer mixture of two ionomers PEDOT:PSS was used as the hole transport and electron blocking layer. The~active layer was composed of a NPs:P3HT:PCBM mixture. Bathophenanthroline (BPhen) played the role of a hole-blocking layer on top of the active layer due to its wide energy gap and high ionization potential~\cite{Chen2002,Ma2016}. Finally,~aluminium film was used as a top electrode, and~the whole structure was encapsulated under a glass. 
A detector without NPs (pure P3HT:PCBM $\sim$3.3~\textmu m thick layer) was also fabricated for comparison. 
The experiment scheme and a photo of one of the detectors are shown in Figure \ \ref{fig4}b. 

\vspace{-6pt}

The P3HT:PCBM blend is one of the most studied and used active materials for bulk heterojunction organic solar cells~\cite{Ameri2013}, which show good efficiency in the visible and UV ranges. In~the X-ray range, the~absorption probability strongly depends on the atomic number $Z$, the~density of the material, and~the incident X-ray photon energy. Therefore, organic polymers exhibit a low X-ray attenuation coefficient (especially in the hard X-ray range) that can be improved by the incorporation of inorganic particles including elements with a high atomic number (high-Z) values~\cite{Thirimanne2018}. 

Note that a sharp rise in the X-ray absorption coefficient, called the X-ray absorption edge, is observed for elements at particular X-ray energies when the incident X-ray energy is equal to that of the binding energy of a core-level electron. For~instance, Ca ($Z$ = 20) and Zn ($Z$ =30) have K edges at 4038.5~eV and 9659~eV, respectively, whereas the more heavy W ($Z$ = 74) has L$_3$-edge at 10,207~eV~\cite{Xray2009}. A~strong increase in absorption beyond the edge can be used for the optimization of detector spectral sensitivity in a required energy range. In~some compounds,~intense absorption, the~so-called ``white line'' (WL), is observed just above the edge. Its existence is common in X-ray absorption spectra of tungsten oxide compounds at the W L$_{2,3}$ edges and is caused by quasilocalized 5d(W) states~\cite{Balerna1991,Kalinko2011}. In~this study, we employed the presence of the WL, located at 10,212~eV at the W L$_3$ edge, to~amplify the signal of detectors containing~tungstates. 

The fabricated hybrid detectors were exposed to monochromatic synchrotron radiation at three selected X-ray photon energies (9500~eV, 10,212~eV, and~20,000~eV) to study the effects of tungstate NPs present in the active layer on the detector response. The~first energy value (9500~eV) was selected between the K edges of Ca and Zn, the~second energy value (10,212~eV) was located at the WL maximum of the tungsten L$_3$ edge, and~the third energy value (20,000~eV) was chosen well above all absorption edges in the two~tungstates. 

When X-ray photons are absorbed in matter, electron--hole pairs are generated via the internal photoelectric effect, followed by an avalanche of secondary generated electrons. In~the detector, the~charge carriers are transported to the electrodes and can be detected as electric current. Note that similar to polymer solar cells~\cite{Cai2010}, our hybrid P3HT:PCBM-based detectors can operate without any external bias~voltage.

Time-dependent X-ray experiments for three detectors were performed at the above-mentioned incident X-ray energies by periodically turning the incident X-rays on and off with a 120~s period. Note that only two X-ray energies (9500~eV and 10,212~eV) were used in the case of the CaWO$_4$-based detector. The~normalized X-ray response on/off signals, i.e.,~the intensity ratio $I_{detector}(E) / I_0(E)$, where $I_0(E)$ corresponds to the incident X-ray intensity measured by the ionization chamber, and I$_{detector}(E)$ is the signal detected by the detector, are shown in Figure\ \ref{fig4}c--e. As~one can see, all three detectors 
demonstrated sensitivity to X-rays in a form close to a square-shaped response, which
differs from the saw-tooth-shaped photocurrent response observed in some organic~\cite{Basirico2016} and hybrid~\cite{Jayawardena2019} X-ray detectors.
Note the different $y$-axis scale in Figure\ \ref{fig4}d. It is difficult to compare absolute values of the photocurrent for different detectors because they are strongly affected by the fabrication quality. Nevertheless, we can compare relative responses at different~energies. 

A response to exposure to X-rays of the pure organic P3HT:PCBM detector is shown in Figure\ \ref{fig4}c.
Because organic polymers are composed of light elements, they exhibit a low X-ray attenuation coefficient and are less sensitive at larger energies. 
As a result, the~photogenerated current is close for X-ray photons at 9500~eV and 10,212~eV
but is smaller at 20,000~eV. 
After X-ray illumination begins, the~top of the current pulse relaxes and drops by about 30--50\% during the first seconds. When the shutter blocs the X-rays, the~dark current becomes slightly negative and relaxes back to the initial value.
The sharp signal at the X-ray shutter switching points is presumably related to the charge trapping/detrapping at material/electrode interfaces and space charge effects~\cite{Haneef2020, Butanovs2021}. 

The X-ray-induced response of both hybrid detectors with nanoparticles (Figure\ \ref{fig4}d,e) was close after sufficient stabilization time and is similar to that found in other hybrid detectors~\cite{Nanayakkara2020,Xiang2021}. The~presence of tungstate nanoparticles significantly (3--4 times) enhanced the response of the detector at the energy of 10,212~eV just above the W L$_3$ edge due to the presence of the strong absorption resonance (``white line'') \cite{Balerna1991,Kalinko2011}. In~the case of the ZnWO$_4$-based detector (Figure\ \ref{fig4}e), an additional contribution came from the absorption caused by the Zn K-edge located at 9659~eV. Note that the X-ray absorption caused by tungsten L$_{1,2,3}$ edges, located at 10,207~eV, 11,544~eV, and~12,100~eV, respectively, contributed as much as up to 20,000~eV. As~a result, 
the response of the ZnWO$_4$-based detector was about two times larger at 20,000~eV than at 9500~eV. 

\begin{figure}[H]
	 
	\includegraphics[width=0.92\textwidth]{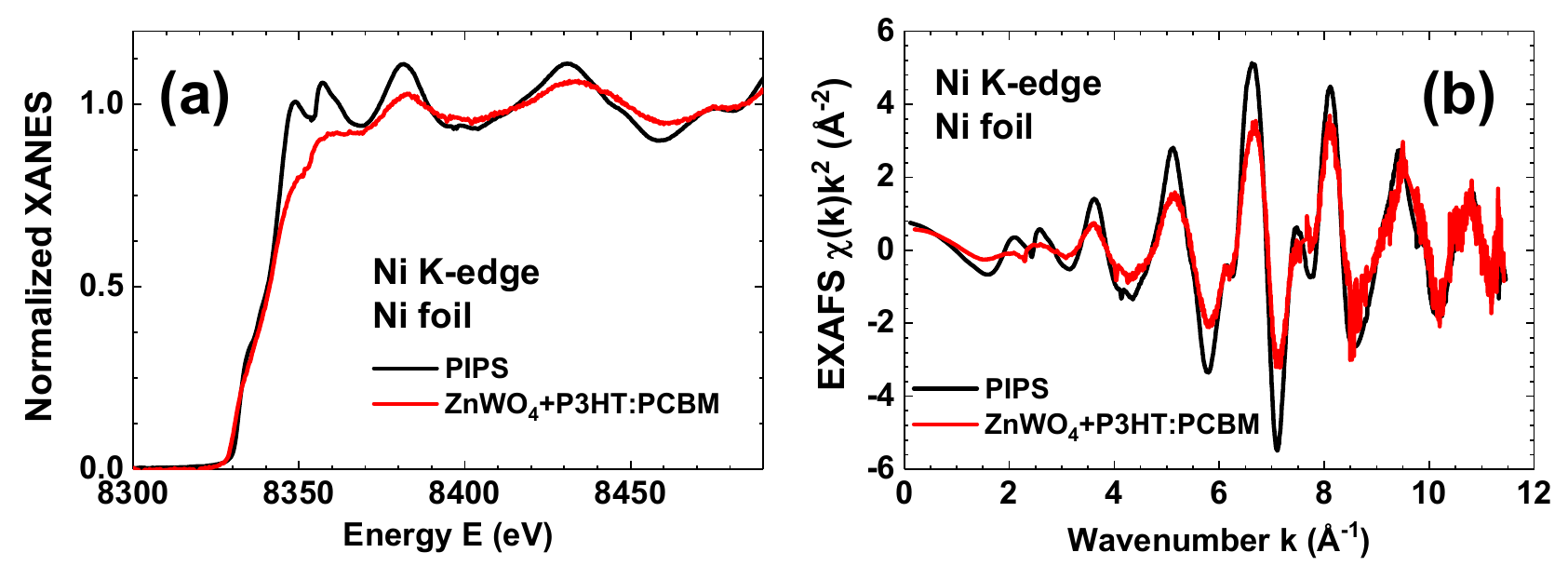}	
	\caption{Comparison of the Ni K-edge (\textbf{a}) XANES $\mu(E)$ and (\textbf{b}) EXAFS spectra $\chi(k)k^2$ of nickel foil measured using novel hybrid and PIPS detectors. $k$ is the wavenumber of the excited~photoelectron. }
	\label{fig5}
\end{figure}

Finally, we used a hybrid detector with ZnWO$_4$ nanoparticles to record the Ni K-edge X-ray absorption spectrum of a Ni foil (Figure\ \ref{fig5}). A PIPS detector was simultaneously used in fluorescence mode for comparison. As~one can see, the~X-ray absorption near-edge structure (XANES) $\mu(E)$ measured in continuous scan mode by the hybrid detector exhibited broadened oscillations; nevertheless, all main spectral features were present. The~observed broadening was caused by the slow response of the hybrid detector compared with PIPS. The~broadening effect decreased in the range of the extended X-ray absorption fine structure (EXAFS) $\chi(k)k^2$ ($k=\sqrt{(2m_e/\hbar^2)(E-E_0)}$, where $m_e$ is the electron mass, $\hbar$ is the Planck constant, and~$E_0$ is the threshold energy, i.e.,~the energy of a free electron with zero momentum), located beyond the absorption edge, due to the frequency of oscillations of the absorption coefficient becoming gradually lower with increasing energy~\cite{Rehr2000}. Note the good agreement between the EXAFS spectra measured by the two detectors at large wave numbers $k$, where the signal is multiplied by the $k^2$ factor. This finding suggests a good dynamic range of the ZnWO$_4$-based hybrid~detector. 

\vspace{-6pt}
\unskip

\section{Conclusions}
Nanocrystalline CaWO$_4$ and ZnWO$_4$ tungstates were studied as promising candidates for use in hybrid organic--inorganic direct-conversion X-ray detectors operating without a bias voltage. The~tungstate nanoparticles (NPs) with a crystallite size of $\sim$43~nm for CaWO$_4$ and $\sim$30~nm for ZnWO$_4$ were synthesized by the hydrothermal method with different~morphologies. 

\textls[-5]{Hybrid organic--inorganic X-ray detectors were fabricated on top of ITO-covered glass and had a sandwich-type structure composed of five ITO/PEDOT:PSS/NPs:P3HT: PCBM/BPhen/Al layers.
A pure organic detector without nanoparticles was used for comparison. 
 Feasibility experiments were performed using monochromatic synchrotron radiation allowing us to conduct the measurements in a wide X-ray energy range (\mbox{9000--20,000~eV}). 
We showed that the presence of tungstate nanoparticles with high-Z elements increases the X-ray attenuation efficiency and, thus, improves the response of the hybrid detector to X-rays compared with those of a pure organic one based on a P3HT:PCBM bulk heterojunction cell.
The use of the developed detector for spectroscopic applications was also demonstrated by recording the Ni K-edge X-ray absorption spectrum of nickel foil. Its well-resolved and extended fine X-ray absorption structure indicates the high dynamic range of the~detector. }

Such hybrid detectors with different AWO$_4$ tungstate nanoparticles can provide a cost-effective solution that can be optimized for a particular energy range by selecting the A-cation type and operating without external~voltage.

%%%%%%%%%%%%%%%%%%%%%%%%%%%%%%%%%%%%%%%%%%
\vspace{6pt} 

%%%%%%%%%%%%%%%%%%%%%%%%%%%%%%%%%%%%%%%%%%
%% optional
%\supplementary{The following supporting information can be downloaded at: \linksupplementary{s1}, Figure S1: title; Table S1: title; Video S1: title.}


%%%%%%%%%%%%%%%%%%%%%%%%%%%%%%%%%%%%%%%%%%
\authorcontributions{Conceptualization, I.P. and K.P.; Data Curation, I.P. and K.P.; Funding Acquisition, A.K. (Aleksandr Kalinko); Investigation, I.P., K.P., A.T., N.R.S., A.K. (Aleksandr Kalinko) and A.K. (Alexei Kuzmin); Methodology, I.P., K.P., A.K. (Aleksandr Kalinko) and A.K. (Alexei Kuzmin); Project Administration, A.K. (Aleksandr Kalinko); Resources, A.K. (Aleksandr Kalinko); Supervision, A.K. (Alexei Kuzmin); Writing---Original Draft, I.P. and A.K. (Alexei Kuzmin); Writing---Review and Editing, I.P., K.P., A.K. (Aleksandr Kalinko) and A.K. (Alexei Kuzmin). 
All authors have read and agreed to the published version of the~manuscript.} 

\funding{The authors are thankful for the financial support from the Latvian Council of Science project No. lzp-2019/1-0071.}

\institutionalreview{Not applicable.}

\informedconsent{Not applicable.}

\dataavailability{The data presented in this study are available on request from the corresponding author. The~data are not publicly available because of ongoing research.} 

\acknowledgments{The experiment at the DESY PETRA-III synchrotron was performed within project No. I-20211105 EC at the Institute of Solid State Physics, University of Latvia, as the Center of Excellence has received funding from the European Union's Horizon 2020 Framework Programme H2020-WIDESPREAD-01-2016-2017-TeamingPhase2 under grant agreement No. 739508, \linebreak project~CAMART2.}

\conflictsofinterest{The authors declare no conflict of~interest.} 

%%%%%%%%%%%%%%%%%%%%%%%%%%%%%%%%%%%%%%%%%%

\abbreviations{Abbreviations}{
The following abbreviations are used in this manuscript:\\

\noindent 
\begin{tabular}{@{}ll}

EXAFS & Extended X-ray absorption fine structure\\
NP & nNanoparticle\\
PEDOT:PSS & Poly(3,4-ethylenedioxythiophene)-poly(styrenesulfonate)\\
P3HT:PCBM & Poly(3-hexylthiophene-2,5-diyl):Phenyl-C61-butyric acid methyl ester\\
RT & Room temperature\\
SEM & Scanning electron microscopy\\
WL & White line\\
XANES & X-ray absorption near edge structure \\
XRD & X-ray diffraction
\end{tabular}
}

%%%%%%%%%%%%%%%%%%%%%%%%%%%%%%%%%%%%%%%%%%
\begin{adjustwidth}{-\extralength}{0cm}

\reftitle{References}

% Please provide either the correct journal abbreviation (e.g. according to the �List of Title Word Abbreviations� http://www.issn.org/services/online-services/access-to-the-ltwa/) or the full name of the journal.


\PublishersNote{}

% If authors have biography, please use the format below
%\section*{Short Biography of Authors}
%\bio
%{\raisebox{-0.35cm}{\includegraphics[width=3.5cm,height=5.3cm,clip,keepaspectratio]{Definitions/author1.pdf}}}
%{\textbf{Firstname Lastname} Biography of first author}
%
%\bio
%{\raisebox{-0.35cm}{\includegraphics[width=3.5cm,height=5.3cm,clip,keepaspectratio]{Definitions/author2.jpg}}}
%{\textbf{Firstname Lastname} Biography of second author}

% For the MDPI journals use author-date citation, please follow the formatting guidelines on http://www.mdpi.com/authors/references
% To cite two works by the same author: \citeauthor{ref-journal-1a} (\citeyear{ref-journal-1a}, \citeyear{ref-journal-1b}). This produces: Whittaker (1967, 1975)
% To cite two works by the same author with specific pages: \citeauthor{ref-journal-3a} (\citeyear{ref-journal-3a}, p. 328; \citeyear{ref-journal-3b}, p.475). This produces: Wong (1999, p. 328; 2000, p. 475)

%%%%%%%%%%%%%%%%%%%%%%%%%%%%%%%%%%%%%%%%%%
%% for journal Sci
%\reviewreports{\\
%Reviewer 1 comments and authors� response\\
%Reviewer 2 comments and authors� response\\
%Reviewer 3 comments and authors� response
%}
%%%%%%%%%%%%%%%%%%%%%%%%%%%%%%%%%%%%%%%%%%
\end{adjustwidth}
\end{document}